\documentclass[%
reprint,
superscriptaddress,
 amsmath,amssymb,
 aps,
]{revtex4-2}

\usepackage[T1]{fontenc}
\usepackage[utf8]{inputenc}
\usepackage{array}
\usepackage{float}
\usepackage{multirow}
\usepackage{amsmath}
\usepackage{amssymb}
\usepackage{cancel}

\usepackage{bbm}
\usepackage{array}
\usepackage{braket}
\usepackage{hyperref}
\usepackage{siunitx}

\usepackage{graphicx}
\usepackage{dcolumn}
\usepackage{bm}
\usepackage{bbm}
\usepackage{siunitx}
\usepackage{hyperref}

\begin{document}
\title{Tunable superconducting flux qubits with long coherence times}
\author{T. \surname{Chang}}
\affiliation{Quantum Nanoelectronics Laboratory, Department of Physics \& Bar-Ilan
Institute of Nanotechnology and Advanced Materials (BINA), 5290002
Ramat-Gan, Israel.}
\author{T. \surname {Cohen}}
\affiliation{Quantum Nanoelectronics Laboratory, Department of Physics \& Bar-Ilan
Institute of Nanotechnology and Advanced Materials (BINA), 5290002
Ramat-Gan, Israel.}
\author{I. \surname{Holzman}}
\affiliation{Quantum Nanoelectronics Laboratory, Department of Physics \& Bar-Ilan
Institute of Nanotechnology and Advanced Materials (BINA), 5290002
Ramat-Gan, Israel.}
\author{G. \surname{Catelani}}
\affiliation{JARA Institute for Quantum Information (PGI-11), Forschungszentrum
J\"{u}lich, 52425 J\"{u}lich, Germany}
\affiliation{Quantum Research Centre, Technology Innovation Institute, Abu Dhabi,
United Arab Emirates}
\author{M. \surname{Stern}}
\affiliation{Quantum Nanoelectronics Laboratory, Department of Physics \& Bar-Ilan
Institute of Nanotechnology and Advanced Materials (BINA), 5290002
Ramat-Gan, Israel.}
\email{michael.stern@biu.ac.il}

\date{15 June 2022}

\newcommand{\hamil}{\mathcal{H}}
\newcommand{\phir}{\Phi_{R}}
\newcommand{\phis}{\Phi_{S}}
\newcommand{\phisr}{\Phi_{S/R}}
\newcommand{\sorr}{S\left(R\right)}
\newcommand{\ixs}{I_{x, S}}
\newcommand{\izs}{I_{z, S}}
\newcommand{\ixr}{I_{x, R}}
\newcommand{\izr}{I_{z, R}}
\newcommand{\ixsr}{I_{x, S/R}}
\newcommand{\izsr}{I_{z, S/R}}
\newcommand{\gfsr}{\Gamma_{S/R}}
\newcommand{\gfs}{\Gamma_{S}}
\newcommand{\gfr}{\Gamma_{R}}
\newcommand{\gfso}{\Gamma_{2nd}}
\newcommand{\gffo}{\Gamma_{1st}}
\newcommand{\micron}{\si{\micro\meter}}
\newcommand{\br}[1]{\Bra\{#1\}}
\newcommand{\kt}[1]{\Ket\{#1\}}
\newcommand{\brkt}[1]{\Braket\{#1\}}

\newcommand{\microsec}{\si{\micro\second}}


\begin{abstract}
In this work, we study a series of tunable flux qubits inductively
coupled to a coplanar waveguide resonator fabricated on a sapphire
substrate. Each qubit includes an asymmetric superconducting quantum
interference device which is controlled by the application
of an external magnetic field and acts as a tunable Josephson junction.
The tunability of the qubits is typically $\pm\SI{3.5}{\giga\hertz}$
around their central gap frequency. The
measured relaxation times are limited by dielectric losses in the
substrate and can attain $T_{1}\sim\SI8{\micro\second}$. The echo
dephasing times are limited by flux noise even at optimal points and
reach $T_{2E}\sim\SI4{\micro\second}$, almost an order of magnitude
longer than state of the art.
\end{abstract}

\maketitle


The superconducting flux qubit is a micron-size superconducting aluminium
loop intersected by several Josephson junctions, among which one is
smaller than others by a factor $\alpha $ \citep{PhysRevB.60.15398, Mooij1999, Chiorescu2003, PhysRevLett.95.257002, PhysRevLett.97.167001, PhysRevB.93.104518}.
When the flux threading the loop is close to half a flux quantum,
this circuit behaves as a two-level system and can exhibit long coherence
times \citep{Bylander2011, PhysRevLett.113.123601}. Thus, it is often
considered as a strategic building block for the physical realization
of superconducting quantum computers \citep{PhysRevApplied.13.034037}.
Yet, a good control of the transition energy of the qubit
at its optimal working point is required to perform
efficient gates on a scalable system.

A good strategy for controlling the qubit transition energy consists
of replacing one of the junction by a superconducting quantum interference
device (SQUID). The
advantage of this approach is that another control parameter is added
to the system: the flux $\Phi_{S}$ threading the loop of the SQUID
controls the critical current of the equivalent junction and therefore
modifies the energy of the flux qubit while keeping it at its optimal
point. This kind of design was implemented for the first time in Ref.
 \citep{PhysRevLett.102.090501}: a symmetric SQUID formed by two identical
Josephson junctions was introduced at the position of the $ \alpha $ -junction
in order to control the gap energy of the qubit. The results of this
experiment were positive in terms of controllability
($\sim\SI{0.7}{\giga\hertz}/\text{m\ensuremath{\Phi_{0}}}$) but
the relaxation and dephasing times of the qubit were severely degraded
($T_{1}\lesssim\SI{1}{\micro\second}$, $T_{2}\sim\SI{10}{\nano\second}$).

These short coherence times are generally attributed to the presence
of flux noise in the SQUID loop \citep{PhysRevLett.118.057702, PhysRevApplied.13.054079}.
For any flux-tunable qubit, this flux noise leads to significant dephasing
whenever the qubit energy is too strongly dependent on the external
flux bias $\Phi_{S}$. A possible way to mitigate this issue consists
of using an asymmetric SQUID formed by two different junctions having
respectively a Josephson energy $(1+d)E_{J}/2$ and $(1-d)E_{J}/2$
with $d\in[0, 1]$. The equivalent Josephson energy
of such a SQUID $E_{J}(\Phi_{S})$ varies according to the following
expression \citep{PhysRevA.76.042319}:

\begin{equation}
E_{J}(\Phi_{s})=E_{J}\sqrt{\frac{\left(1+d^{2}\right)+\left(1-d^{2}\right)\text{cos\ensuremath{\left(\frac{\Phi_{S}}{\varphi_{0}}\right)}}}{2}}
\end{equation}
where $\varphi_{0}=\hbar/2e $.
For a given value of $d $,  the function $E_{J}(\Phi_{S})$ ranges
between $ dE_{J}$ and $E_{J}$ and consequently the dependence of the
qubit energy on $\Phi_{S}$ is strongly reduced as $d $ approches
1. This technique has been demonstrated recently for tuning transmon
qubits while keeping good coherence properties \citep{PhysRevB.87.220505, PhysRevApplied.8.044003}.

In this work, we follow the same strategy for controlling the gap
of superconducting flux qubits. We replace one of the unitary junctions
of the flux qubit by an assymmetric SQUID and study the controllability
of the qubits and their coherence properties. The tunability of the
qubits is $\pm\SI{3.5}{\giga\hertz}$ around their central frequency. The intrinsic relaxation rates can be
as low as $ \Gamma_{int}\sim\SI{130}{\kilo\hertz}$ ($T_{1}\sim\SI 8{\micro\second}$)
while the pure echo dephasing rates at optimal points are typically $ \Gamma_{\varphi E}\sim260\pm\SI{90}{\kilo\hertz}$ $(T_{2E}^{\varphi}\sim4\:\mu s )$. These decoherence rates are much
smaller than the state of the art for tunable flux qubits \citep{PhysRevLett.102.090501, PhysRevB.84.014525, Schwarz-2013}.
We show that these decoherence rates are mostly limited by flux noise,
even at optimal points.

\begin{figure*}[t]
\centering{}\includegraphics[width=1\textwidth]{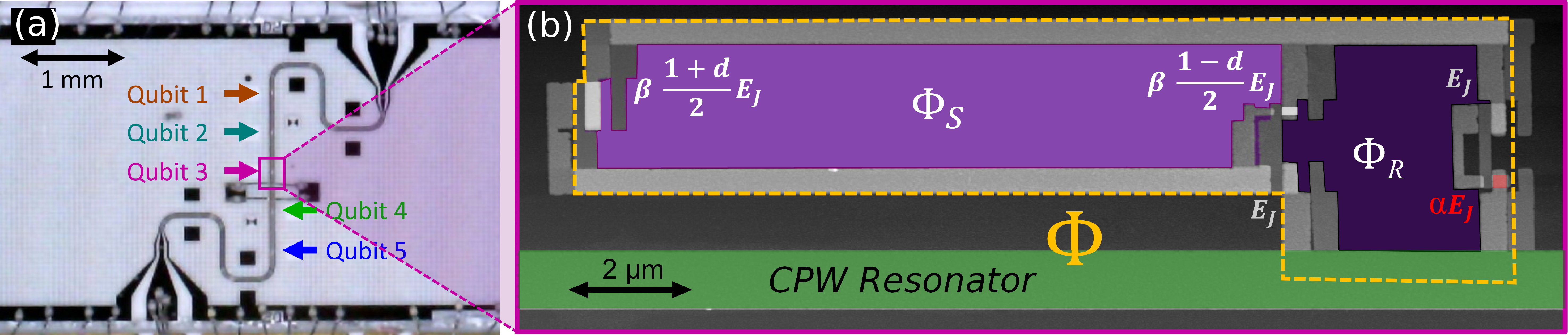}\caption{(a) Microscope picture of the sample, showing the coplanar waveguide
resonator inductively coupled to five tunable flux qubits labelled
according to their position on the resonator. The resonator is fabricated
on a sapphire wafer by evaporation of a 150 nm- thick aluminium layer
and UV lithography. (b) Colored atomic force micrograph of flux qubit
3. The qubit is galvanically coupled to the central conductor of the
resonator (colored in green). It consists of two loops: The surface
of the main loop is $ S_{main}^{(3)}=\protect\SI{43.71}{\micro\meter\squared}$
while the surface of the SQUID loop is $ S_{squid}=\protect\SI{30.82}{\micro\meter\squared}$
giving a ratio $ \zeta^{(3)}=0.705 $. In our experimental setup, the
magnetic field is applied uniformly such that the flux threading the
SQUID loop is $ \Phi_{S}=\zeta\Phi $ and $ \Phi_{R}=(1-\zeta)\Phi $.}
\end{figure*}

The sample studied in this work is presented in Fig. 1a. It is fabricated
on a sapphire chip and contains a $\lambda/2 $ aluminium coplanar
waveguide (CPW) resonator with two symmetric ports for microwave transmission
measurements. The resonator has a first resonant mode at $\omega_{r}/2\pi\sim\SI{10.23}{\giga\hertz}$
with quality factor $Q\sim3500 $. Five tunable flux qubits labelled
according to their position on the resonator $i=\{1,..., 5\}$ are
galvanically coupled to the CPW resonator with coupling constant $\sim\SI{120}{\mega\hertz}$. 

The qubits are fabricated by double angle-evaporation of Al--AlOx--Al
using a tri-layer CSAR-Ge-MAA process \citep{PhysRevLett.113.123601}.
The tri-layer is patterned by electron-beam lithography, developed
and etched by a reactive ion etcher in order to form a suspended germanium
mask. This mask is robust and evacuates efficiently the charges
during e-beam lithography, and thus provides a good precision and
reproducibility of the junction sizes. Before aluminium
evaporation, an ion milling step etches the oxide layer from the central
conductor of the resonator in order to connect it galvanically to
the qubit. A first  $\SI{25}{\nano\meter}$ layer of aluminium is evaporated at 
$\SI{-50}{\celsius}$ in a direction 
of $\SI{-25}{\degree}$ relative to the sample axis. This
step is followed by dynamic oxidation of $\mathrm{O_{2}/Ar}$ (15\%-85\%)
at $P=20\:\mathrm{\text{\ensuremath{\mu}bar}}$ for a duration of
30 minutes. A second layer of 30 nm of aluminium is then evaporated
with the opposite angle ($\SI{+25}{\degree}$) at a temperature of
$\sim\SI{7}{\celsius}$. The low temperature
enables us to reduce the grain size of aluminium and to better control
the dimensions and oxidation properties of our junctions. Before cooling
down the sample, we performed room temperature measurements of reference
junctions. An histogram of these measurements is given in Supplementary
Materials \citep{Suppl}. 

The low temperature measurements of the qubits are performed in a
cryogen-free dilution refrigerator at a temperature of 14 mK. The
input line is attenuated at low temperature to minimize
thermal noise and filtered with homemade impedance-matched radiation-absorbing
filters. The readout output line includes a band-pass filter, a double
circulator and a cryogenic amplifier. Qubit state manipulations
are performed by injecting in the input line of the resonator microwave
pulses at the frequency of the qubit $\omega_{01}$, followed by a
readout pulse at $\omega_{r}$ whose amplitude and phase yield the
qubit excited state probability. The sample is glued on a printed circuit board 
and embedded into a
superconducting coil that is used to provide magnetic flux biases
to the qubits. In order to isolate the device from surrounding magnetic
noise, the system is magnetically shielded with a Cryoperm box surrounding
a superconducting enclosure.

\begin{figure*}
\centering{}\includegraphics[width=1\textwidth]{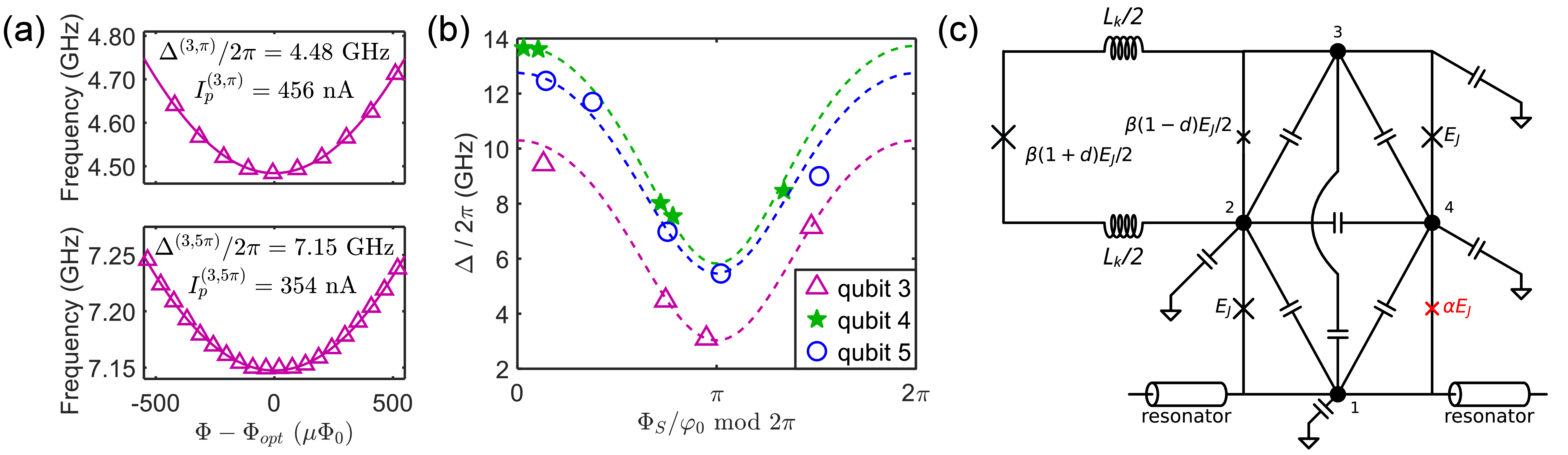}\caption{(a) Measured qubit frequency of qubit 3 versus $\Phi $ (magenta triangles)
and fit (magenta curve) yielding the qubit parameters $\Delta $ and
$I_{p}${\scriptsize{} }at optimal points $\pi $ and $5\pi $. (b)
Measured gaps of qubit 3 (magenta triangles),  4 (green stars) and
5 (blue circles) versus $\Phi_{s}/\varphi_{0}$. The dashed curves
are calculated for arbitrary values of $\Phi_{S}$ according to the
model illustrated in (c) with fitting parameters $E_{J}$, $\alpha $ 
and $d $. (c) Flux qubit model including the kinetic inductance $L_{k}$
and geometric capacitances. The island 1 is connected
galvanically to the central conductor of the resonator.}
\end{figure*}

Fig. 1b presents a colored atomic force micrograph (AFM) of one of
these qubits. The circuit consists of two loops. The main 
loop indicated by a yellow dashed line is intersected by two identical
Josephson junctions of Josephson energy $E_{J}$ and one smaller junction
colored in red of Josephson energy $\alpha E_{J}$. The SQUID loop
is intersected by two additionnal Josephson junctions of Josephson
energy $\beta(1+d)E_{J}/2 $ and $\beta(1-d)E_{J}/2 $. Its surface
is smaller than the main loop by a factor $\zeta\sim0.7 $. 

The inductive energy of the circuit exhibits two local minima which
correspond to a persistent current $I_{P}$ flowing clockwise or anticlockwise
in the main loop. These two minima become degenerate when the flux
threading the main loop $\Phi=\Phi_{opt}$ is such that
\begin{equation}
\frac{\Phi_{opt}}{\varphi_{0}}-\frac{\Phi_{S}}{2\varphi_{0}}+\delta\varphi=k\pi
\end{equation}
 with $k=\pm1,\pm3,\pm5...$ and $\tan\delta\varphi=d\tan\left[\frac{\Phi_{s}}{2\varphi_{0}}\right]$.
At these \emph{optimal} points, the two quantum states hybridise into symmetric
and antisymmetric superpositions and give rise to an energy splitting
$\hbar\Delta $ called the flux-qubit gap. In our experimental setup,
the magnetic field is applied uniformly such that the flux threading
the SQUID loop is $\Phi_{S}=\zeta\Phi $. One can therefore solve
Eq. 2 and get the values of the fluxes $\Phi_{opt}$ and $\Phi_{S}$
at each optimal point. For different values of $k $,  the value of
the effective Josephson energy of the SQUID changes according to Eq.
1 and consequently, the value of the gap of each qubit depends on
$k $. In the following, the gap of qubit $i $ at each optimal point
will be denoted as $\Delta^{(i, k\pi)}$ and its associated persistent
current as $I_{P}^{(i, k\pi)}$. 

Fig. 2a shows the frequency dependence of qubit $3 $ on $\Phi $ 
around the $\pi $ and $5\pi $ optimal points. The transition frequency
of the qubit around each optimal point follows $\omega_{01}=\sqrt{\Delta^{2}+\varepsilon^{2}}$
with $\varepsilon=2I_{P}\left(\Phi-\Phi_{opt}\right)/\hbar $,  yielding
$\Delta^{(3,\pi)}/2\pi=\SI{4.48}{\giga\hertz}$, $I_{P}^{(3,\pi)}=\SI{456}{\nano\ampere}$
and $\Delta^{(3, 5\pi)}/2\pi=\SI{7.15}{\giga\hertz}$, $I_{P}^{(3, 5\pi)}=\SI{354}{\nano\ampere}$.
We repeat this procedure for the five qubits at their respective optimal
points $\pm\pi,\pm3\pi,\pm5\pi,\pm7\pi,\pm9\pi $ (See \citep{Suppl}). 

Fig. 2b presents the gaps of qubits 3, 4 and 5 versus $\Phi_{s}/\varphi_{0}$.
This data together with the persistent currents obtained for each
optimal point \citep{Suppl} enables us to fit parameters of the model
shown in Fig. 2c. In this model, the qubit consists of two superconducting
loops intersected by five Josephson junctions. Each Josephson junction
is characterized by its Josephson energy $E_{J}$ and its bare capacitance
energy $E_{C}=e^{2}/2C_{J}$. The junctions divide the loops into
four superconducting islands. Each island is capacitively coupled
to its surrounding by geometric capacitances. 
These geometric capacitances are calculated using the electrostatic module of COMSOL (see \citep{Suppl}). They reduce the gaps of the qubits by approximately $\sim\SI{1}{\giga\hertz}$ but barely
modify their persistent currents. It is also neccessary
to take into account the kinetic inductance of the SQUID loop in order
to match the parameters of the model with the experimental results.
The kinetic inductance is estimated by measuring the resistance of
evaporated aluminium wires at low temperature and is added in our
model as a renormalization of the large Josephson junction of the
SQUID \citep{Suppl}. We summarize the results of the fits in Table
1. These values are in good agreement with the measured values of
$\alpha, d $ and $E_{J}$ extracted from room temperature resistance
measurements (see \citep{Suppl}). 

\begin{table}
\begin{tabular}{|c||c|c|c|c|c|c|}
\hline 
{\footnotesize{}Qubit \#} & {\footnotesize{}$E_{J}$ (GHz)} & {\footnotesize{}$E_{J}/E_{C}$} & {\footnotesize{}$\alpha$} & {\footnotesize{}$d$} & {\footnotesize{}$\zeta$} & {\footnotesize{}$\sqrt{A_{R}A_{S}}$ ($\mu\varPhi_{0})$}\tabularnewline
\hline 
{\footnotesize{}1} & {\footnotesize{}550} & {\footnotesize{}318} & {\footnotesize{}0.429} & {\footnotesize{}0.759} & {\footnotesize{}0.696} & {\footnotesize{}2.2}\tabularnewline
\hline 
{\footnotesize{}2} & {\footnotesize{}558} & {\footnotesize{}323} & {\footnotesize{}0.426} & {\footnotesize{}0.715} & {\footnotesize{}0.687} & {\footnotesize{}2.6}\tabularnewline
\hline 
{\footnotesize{}3} & {\footnotesize{}559} & {\footnotesize{}323} & {\footnotesize{}0.442} & {\footnotesize{}0.711} & {\footnotesize{}0.705} & {\footnotesize{}2.3}\tabularnewline
\hline 
{\footnotesize{}4} & {\footnotesize{}518} & {\footnotesize{}300} & {\footnotesize{}0.421} & {\footnotesize{}0.707} & {\footnotesize{}0.677} & {\footnotesize{}2.3}\tabularnewline
\hline 
{\footnotesize{}5} & {\footnotesize{}563} & {\footnotesize{}326} & {\footnotesize{}0.426} & {\footnotesize{}0.740} & {\footnotesize{}0.714} & {\footnotesize{}3.0}\tabularnewline
\hline 
\end{tabular}

\caption{Parameters of the qubits. The charging energy is fixed at
$E_{C}=\SI{1.73}{\giga\hertz}$ assuming a specific capacitance of the
junction $C/A=\protect\SI{100}{\femto\farad\per\micro\meter\squared}$.
The ratio $\zeta $ is measured by AFM.
The value $\beta=2.11 $ is taken according to room temperature measurements.
The inductance $L_{k}=\protect\SI{72.5}{\pico\henry}$
is determined according to the resistance measurement of wires at
low temperatures. The value of $E_{J}$, $d $ and $\alpha $ are obtained
by fit with the model shown in Fig. 2c. The amplitude of the flux
noise $\sqrt{A_{R}A_{S}}$ is extracted for each qubit from the dependence
of $\Gamma_{\varphi E}$ versus $\varepsilon $.}
\end{table}

\begin{figure*}
\centering{}\includegraphics[width=1\textwidth]{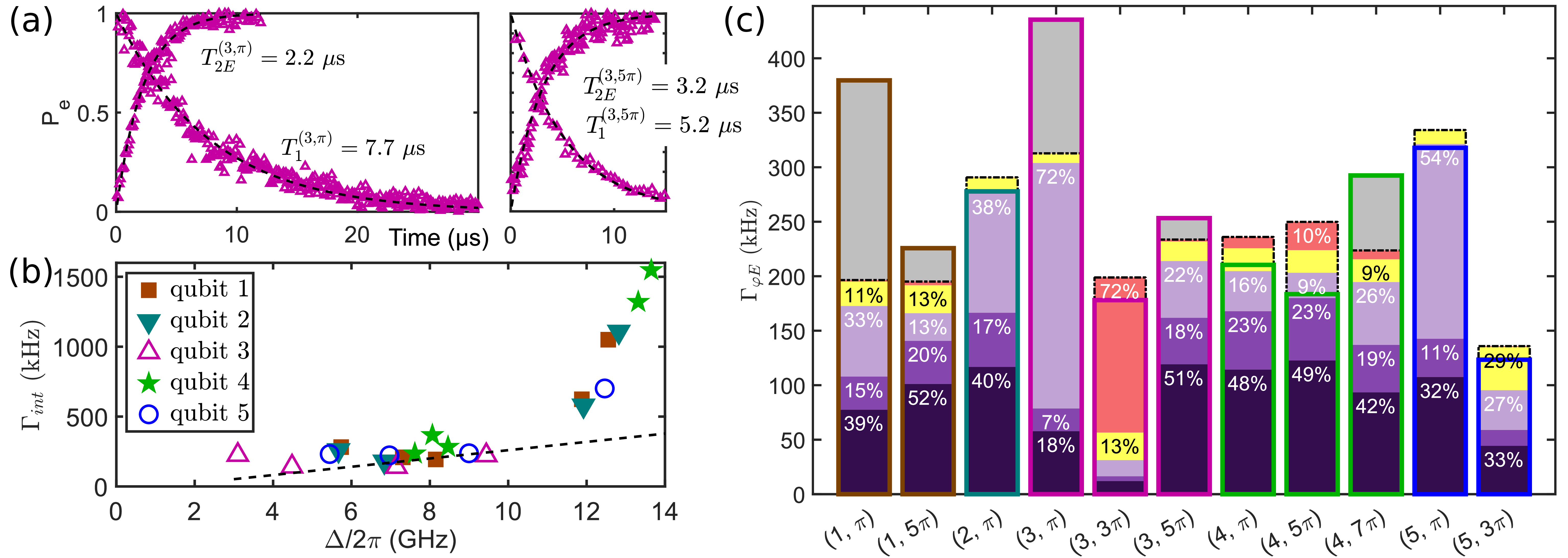}\caption{(a) Energy relaxation and spin-echo measurements of qubit 3 at optimal
points $\pi $ and $5\pi $. (b) Intrinsic relaxation rates $\Gamma_{int}$
versus $\Delta $. The dashed line corresponds to calculated dielectric
losses in the substrate assuming $\tan\delta=5\times10^{-5}$. (c)
Stacked bar chart showing the measured pure echo dephasing rates $\Gamma^{(i,k\pi)}_{\varphi E}$. The calculated contributions
to dephasing are represented in different colors: flux noise (purple), critical current noise $\Gamma_{\alpha}$
(yellow) and photon noise $\Gamma_{phot}$ (red). The flux noise dephasing can be separated into first order flux noise in the
qubit loop $\protect\gfr $ (dark purple), first order flux noise in
the SQUID loop $\protect\gfs $ (purple) and second order effects $\protect\gfso $ 
(light purple). The total $\Gamma_{tot}=\sqrt{\protect\gfr^{2}+\protect\gfs^{2}+\protect\gfso^{2}+\Gamma_{\alpha}^{2}+\Gamma_{phot}^{2}}$
is represented as a dot-dashed black segment. For each contribution
$X\in\{R, S, 2nd,\alpha, phot\}$, the percentage indicated in the relevant
colored stack represents $\Gamma_{X}^{2}/\Gamma_{tot}^{2}$. }
\end{figure*}

The change of the gap modifies the relaxation rate between the two
qubit levels. For illustration, we represent in Fig. 3a the energy
relaxation decay of qubit 3 at two different optimal points. The decay
is exponential in both cases but the relaxation times are different,
namely $T_{1}^{(3,\pi)}=\SI{7.7}{\micro\second}$ and $T_{1}^{(3, 5\pi)}=\SI{5.2}{\micro\second}$.
Several mechanisms may give rise to such a phenomenon; among them, Purcell effect.
The Purcell rate $\Gamma_{P}$ is quantitatively determined by measuring
the qubit Rabi frequency $\Omega_{R}$ for a given microwave power
$P_{in}$ at the resonator input \citep{PhysRevLett.113.123601}.
This enables us to analyze and compare the intrinsic relaxation rates
defined as $\Gamma_{int}=\Gamma_{1}-\Gamma_{P}$ of all the qubits
at various optimal points. Such an analysis shows that qubit 3 has
approximately the same intrinsic relaxation rate at optimal points
$\pi $ and $5\pi $. 

Fig. 3b unveils a
general behavior of intrinsic relaxation rates versus frequency. Previous
measurements of flux qubits \citep{PhysRevLett.113.123601} identified
dielectric losses in the substrate as a major contributor to relaxation
at low temperature \citep{OConnell2008};
using the approach of Ref. \citep{PhysRevLett.113.123601} and a loss
tangent of $5\times10^{-5}$, we obtain the dashed
line in Fig. 3b. Clearly dielectric losses account for most of the relaxation at intermediate frequencies but cannot explain
the increased relaxation rates at high frequencies when the flux in
the SQUID $ \Phi_{S}/\varphi_{0}$ is close to $ 2\pi $. A second source
of losses could be quasiparticle tunneling \citep{PhysRevLett.106.077002, PhysRevB.84.064517, 10.21468/SciPostPhysLectNotes.31}.
A single quasiparticle trapped in one of the large qubit islands would
lead to a relaxation rate larger than what is observed, at least at
low frequencies, as well as to non-exponential decay due to fluctuations
in the number of trapped quasiparticles \citep{Pop2014, Gustavsson2016}.
Alternatively, quasiparticles can reach the qubits from the CPW resonator. 
However, a relatively high normalized quasiparticle density, corresponding
to an effective quasiparticle temperature of order 150mK, would be
needed to explain a decay rate of the order of
tens of \si{\kilo\hertz}. Therefore we conclude that quasiparticles do not significantly 
account for relaxation and cannot explain explain the residual decay rate observed at high frequencies.

At their respective optimal points, the amplitude of the spin-echo
signal shown in Fig. 3a decays with pure dephasing times $T_{\varphi E}^{(3,\pi)}=\SI{2.6}{\micro\second}$
and $T_{\varphi E}^{(3, 5\pi)}=\SI{4.6}{\micro\second}$. In Fig. 3c,
we present a stacked bar chart showing the measured pure echo dephasing
rates at various optimal points and the different contributions of
flux noise, critical current noise, charge noise, and photon noise.
The dephasing due to photon noise (represented in red) has been estimated
by measuring the dispersive shifts at optimal points and by estimating
the number of thermal photons in the resonator \citep{PhysRevLett.95.257002}. It
is the dominant dephasing mechanism for $ (3, 3\pi)$ since at that point the qubit gap happens to be very close to the resonator transition. The contribution of charge noise is strongly reduced by the ratio $E_{J}/E_{C}\sim300$ and was found to be always completely negligible ($<\SI{1}{\kilo\hertz}$). 
We also considered critical current fluctuations
in the $\alpha $ junction assuming $S_{I_{\alpha}}\left(\omega\right)=A_{I_{\alpha}}^{2}/\left|\omega\right|$,
with $A_{I_{0}}\sim0.1\ \text{pA}$ \citep{PhysRevLett.93.077003, Eroms2006},
and found an approximately constant contribution of $\sim\SI{70}{\kilo\hertz}$
(represented in yellow in Fig. 3c). The flux noise shown in purple
represents the main source of dephasing of the qubits even at optimal
points.

Away from their optimal points, the decoherence of flux qubits is
known to be governed by flux noise \citep{PhysRevLett.97.167001, Bylander2011, PhysRevLett.113.123601}.
The flux noise power spectrum $S_{\Phi}(\omega)=A_{\Phi}^{2}/\left|\omega\right|$
implies that the pure echo dephasing rate is given by $\Gamma_{\varphi E}^{\Phi}=A_{\Phi}\sqrt{\ln2}\,\left|\partial\omega_{01}/\partial\Phi\right|$
with $\left|\partial\omega_{01}/\partial\Phi\right|\simeq2\, I_{P}\left|\varepsilon\right|/\hbar\Delta $ 
 \citep{PhysRevB.72.134519}. At the optimal points, $\varepsilon=0 $ 
and thus this decoherence mechanism should be cancelled. Yet, contrary
to \emph{standard} flux qubits, our design contains two independent
degrees of freedom ($\Phi_{S}$, $\Phi_{R}$) \citep{PhysRevB.84.014525}.
These degrees of freedom add $ \sigma_{z}$ components in the Hamiltonian
of the system, namely $ \hamil=\hbar\frac{\Delta}{2}\sigma_{z}+\left(\izs\delta\phis+\izr\delta\phir\right)\sigma_{z}$
(See \citep{Suppl}). Thus, even at optimal point where $ \izs\zeta+\izr\left(1-\zeta\right)=0 $,  $\partial\omega_{01}/\partial\phisr\neq0 $ 
will give first order contributions to dephasing:
\begin{equation}
\gfsr=2\sqrt{\ln2}\frac{\izsr A_{S/R}}{\hbar}
\end{equation}

For each qubit, we measure $\Gamma_{\varphi E}$ versus $\varepsilon $ 
and extract the apparent flux noise amplitude $A_{\Phi}$ around each
optimal point. The amplitudes $A_{S/R}$ of the flux noise in the different
loops can be directly extracted from $A_{\Phi}$
and from the ratio $\gamma=\sqrt{P_{S}/P_{R}}$ where $P_{S}$ and $P_{R}$
are the perimeters of the two loops \citep{PhysRevApplied.13.054079}. As
expected, we find that $ A_{S}$ and $ A_{R}$ do not change significantly
for the different optimal points of a given qubit and thus $\sqrt{A_{S}A_{R}}$
is a good indicator of flux noise in each qubit (See Table 1). A more
rigorous derivation of flux noise contributions including second
order effects is given in \citep{Suppl}. We find that such effects can
be also significant as shown in Fig. 3c.

In conclusion, we have shown that it is possible to control the gap
of flux qubits by using an asymmetric SQUID. This method mitigates
the decoherence due to flux noise in the SQUID loop while keeping
a tunability range of $\pm\SI{3.5}{\giga\hertz}$. It should be possible
to improve further the coherence properties of the qubits by reducing
the persistent currents down to 200 nA and by exchanging the locations
of the small and large junctions of the SQUID. This exchange will
further reduce the tunability of the qubit to the level of $\pm\SI{500}{\mega\hertz}$ and thus decrease the pure dephasing rates related to the presence
of the SQUID. According to our simulations, the dephasing rate due
to flux noise should then be comprised between 15 and $\SI{100}{\kilo\hertz}$.

\begin{acknowledgments}
This research was supported by the Israeli Science Foundation under
grant numbers 426/15, 898/19 and 963/19. G.C. acknowledges support
by the German Federal Ministry of Education and Research (BMBF), funding
program\emph{ Quantum technologies - from basic research to market},
project QSolid (Grant No. 13N16149).

T. Chang and T. Cohen contributed equally to this work.

\end{acknowledgments}
\begin{widetext}

\global\long\def\br#1{\Bra{#1}}%
\global\long\def\kt#1{\Ket{#1}}%
\global\long\def\brkt#1{\Braket{#1}}%
\global\long\def\units#1{\text{#1}}%
\global\long\def\micron{\si{\micro\meter}}%
\global\long\def\microsec{\si{\micro\second}}%
\global\long\def\identity{\mathbbm{1}}%
\global\long\def\cgeom{\mathbf{C_{geom}}}%
\global\long\def\cjun{\mathbf{C_{J}}}%
\global\long\def\cmat{\mathbf{C}}%
\global\long\def\hamil{\mathcal{H}}%
\global\long\def\phir{\Phi_{R}}%
\global\long\def\phis{\Phi_{S}}%
\global\long\def\phisr{\Phi_{S/R}}%
\global\long\def\real{\mathcal{R}}%
\global\long\def\imag{\mathcal{I}}%
\global\long\def\bigo{\mathcal{O}}%
\global\long\def\sorr{S\left(R\right)}%
\global\long\def\ixs{I_{x,S}}%
\global\long\def\izs{I_{z,S}}%
\global\long\def\ixr{I_{x,R}}%
\global\long\def\izr{I_{z,R}}%
\global\long\def\ixsr{I_{x,S/R}}%
\global\long\def\izsr{I_{z,S/R}}%
\global\long\def\gfso{\Gamma_{2nd}}%
\global\long\def\gffo{\Gamma_{1st}}%

\part*{Supplementary Materials}
\section{Tunable flux qubit model}

In the model shown in Fig. 2c, the circuit consists of two superconducting
loops intersected by five Josephson junctions. Each Josephson junction
is characterized by its Josephson energy $E_{J}$ and its bare capacitance
energy $E_{C}=e^{2}/2C_{J}$. The junctions divide the loops into
four superconducting islands. The island 1 is galvanically connected
to the coplanar waveguide resonator. Each island is capacitively coupled
to its surrounding by geometric capacitances. We take also into account
the kinetic inductance of the SQUID loop. This kinetic inductance
can be considered in a first approximation as a renormalization of
the large Josephson junction of the SQUID as will be shown hereinbelow.

\subsection{Treating the kinetic inductance as a renormalization\label{subsec:kind}}

The kinetic inductance of the SQUID loop is represented in the circuit
of Fig. 2c as a an inductor of inductance $L_{k}$ in series with
the large SQUID junction. To find the renormalized parameters for
this junction, we treat it as a linear element with admittance 

\begin{equation}
Y_{J}(\omega)=\frac{1}{i\omega L_{J}}+i\omega C_{J}=\frac{1-(\omega/\omega_{p})^{2}}{i\omega L_{J}}\label{eq:YJ1}
\end{equation}
with $\omega_{p}=1/\sqrt{L_{J}C_{J}}$ the plasma frequency of the
junction. Adding the kinetic inductance in series, we find for the
total admittance:
\begin{equation}
Y_{t}(\omega)=\left(i\omega L_{k}+\frac{i\omega L_{J}}{1-(\omega/\omega_{p})^{2}}\right)^{-1}=\frac{1-(\omega/\omega_{p})^{2}}{i\omega L_{J}+i\omega L_{k}(1-(\omega/\omega_{p})^{2})}\label{eq:Yt1}
\end{equation}
For frequencies small compared to the plasma frequency, the expansion
of this formula at first order in $(\omega/\omega_{p})^{2}$ gives
\begin{equation}
Y_{t}(\omega)\simeq\frac{1-(\omega/\omega_{p})^{2}\frac{L_{J}}{L_{J}+L_{k}}}{i\omega(L_{J}+L_{k})}\label{eq:Ytapp1}
\end{equation}
Comparing the right-hand sides of Eqs.(\ref{eq:YJ1}) and (\ref{eq:Ytapp1}),
we see that the latter can be obtained from former upon the replacements
\begin{align*}
L_{J} & \to L_{J}+L_{k}=L_{J}(1+\eta)\\
C_{J} & \to C_{J}\left(\frac{L_{J}}{L_{J}+L_{k}}\right)^{2}=\frac{C_{J}}{(1+\eta)^{2}}
\end{align*}
where we introduced the dimensionless parameter $\eta=L_{k}/L_{J}$.
In our model we use these replacement to take into account the kinetic
inductance of the SQUID loop.

\subsection{Potential Energy}

The potential energy of the circuit corresponds to the inductive energy
of the junctions and can be written as

\[
U=-E_{J}\cos\varphi_{12}-\beta\frac{1+d}{2\left(1+\eta\right)}E_{J}\cos\varphi_{23}-\beta\frac{1-d}{2}E_{J}\cos\varphi'_{23}-E_{J}\cos\varphi_{23}-\alpha E_{J}\cos\varphi_{41}
\]
where $\varphi_{j,k}$ denotes the phase difference $\varphi_{k}-\varphi_{j}$
between islands $j$ and $k$. 

Introducing

\begin{align*}
\bar{\beta} & =\beta\left(\frac{1+d}{2\left(1+\eta\right)}+\frac{1-d}{2}\right)\\
\bar{d} & =\frac{\left(2+\eta\right)d-\eta}{-\eta d+\left(2+\eta\right)}
\end{align*}

enables us to write the potential energy under the form
\begin{equation}
U=-E_{J}\cos\varphi_{12}-\bar{\beta}\frac{1+\bar{d}}{2}E_{J}\cos\varphi_{23}-\bar{\beta}\frac{1-\bar{d}}{2}E_{J}\cos\varphi'_{23}-E_{J}\cos\varphi_{23}-\alpha E_{J}\cos\varphi_{41}\label{eq:ham-pot}
\end{equation}

Faraday law implies that 
\begin{align}
\varphi_{41} & =\frac{\Phi}{\varphi_{0}}-\sum_{j=1}^{3}\varphi_{j,j+1}\\
\varphi'_{23} & =\varphi_{23}-\frac{\Phi_{S}}{\varphi_{0}}
\end{align}
where $\Phi$ is the flux threading the qubit loop, $\Phi_{S}$ is
the flux threading the SQUID loop and $\varphi_{0}=\hbar/2e$. 

One can write
\begin{align}
 & \bar{\beta}\frac{1+\bar{d}}{2}E_{J}\cos\varphi_{23}+\bar{\beta}\frac{1-\bar{d}}{2}E_{J}\cos\varphi'_{23}\nonumber \\
 & =\bar{\beta}E_{J}\sqrt{\frac{\left(1+\bar{d}^{2}\right)+\left(1-\bar{d}^{2}\right)\text{cos\ensuremath{\left(\frac{\Phi_{s}}{\varphi_{0}}\right)}}}{2}}\cos\left(\varphi_{23}-\frac{\Phi_{s}}{2\varphi_{0}}+\arctan\left(\bar{d}\tan\left[\frac{\Phi_{s}}{2\varphi_{0}}\right]\right)\right)
\end{align}

At optimal point, the sum of the phases across all junctions (including
the SQUID effective junction) should be a odd multiple of $\pi$.
We thus obtain the condition for optimal points given by Eq.2 of the
main text

\[
\boxed{\frac{\Phi_{opt}}{\varphi_{0}}-\frac{\Phi_{s}}{2\varphi_{0}}+\overline{\delta\varphi}=k\pi}
\]
 with $k=\pm1,\pm3,\pm5...$ and $\tan\overline{\delta\varphi}=\bar{d}\tan\left[\frac{\Phi_{s}}{2\varphi_{0}}\right]$.

\subsection{Kinetic Energy}

The kinetic energy $K$ of the system is the sum of the capacitive
energies of the circuit 

\begin{equation}
K=\frac{1}{2}\sum_{i\neq j}C_{ij}\left(V_{j}-V_{i}\right)^{2}+\frac{1}{2}C_{J}\left(\left(V_{1}-V_{2}\right)^{2}+\beta'\left(V_{2}-V_{3}\right)^{2}+\left(V_{3}-V_{4}\right)^{2}+\alpha\left(V_{4}-V_{1}\right)^{2}\right)
\end{equation}
where $C_{ij}$ is the capacitance between islands $i$ and $j$ and
$\beta'=\beta\left(\frac{1+d}{2\left(1+\eta\right)^{2}}+\frac{1-d}{2}\right)$
according to \ref{subsec:kind}. It is a quadratic form of the island
voltages $V_{i}$ and can thus be written as 

\begin{equation}
K=\frac{1}{2}\mathbf{V}^{T}\cmat\mathbf{V}
\end{equation} 
where $\mathbf{V}^{T}=\left(\begin{array}{cccc}
V_{1} & ,V_{2} & ,V_{3}, & V_{4}\end{array}\right)$ and $\cmat$ is a $4\times4$ matrix which we will refer in the following
as the capacitance matrix. The matrix $\cmat$ can be written as the
sum of the Josephson capacitance matrix $\cjun$ and the geometric
capacitance matrix $\cgeom$: 

\begin{equation}
\cmat=\cjun+\cgeom
\end{equation}

where
\begin{equation}
\cjun=C_{J}\left(\begin{array}{cccc}
1+\alpha & -1 & 0 & -\alpha\\
-1 & 1+\beta' & -\beta' & 0\\
0 & -\beta' & 1+\beta' & -1\\
-\alpha & 0 & -1 & 1+\alpha
\end{array}\right)
\end{equation}

and
\begin{equation}
\cgeom=\left(\begin{array}{cccc}
C_{10}+\sum_{j\neq1}C_{1j} & -C_{12} & -C_{13} & -C_{14}\\
-C_{21} & C_{20}+\sum_{j\neq2}C_{2j} & -C_{23} & -C_{24}\\
-C_{31} & -C_{32} & C_{30}+\sum_{j\neq3}C_{3j} & -C_{34}\\
-C_{41} & -C_{42} & -C_{43} & C_{40}+\sum_{j\neq4}C_{4j}
\end{array}\right)
\end{equation}

We determined the geometric capacitance matrix $\cgeom$ using an
electrostatic simulator (COMSOL) and find 
\[
\cgeom=\left(\begin{array}{cccc}
25.0000 & -0.8329 & -0.4047 & -0.1476\\
-0.8329 & 1.8569 & -0.6369 & -0.0063\\
-0.4047 & -0.6369 & 1.9831 & -0.1700\\
-0.1476 & -0.0063 & -0.1700 & 0.3541
\end{array}\right)\qquad\mathrm{fF}
\]

\subsection{Pseudo-Hamiltonian}

We write the Hamiltonian using Legendre transformation and diagonalize
it. We obtain the spectrum of the flux qubit by subtracting the energy
of the first excited state $\kt 1$ from the energy of the ground
state $\kt 0$. Close to the optimal points, the system behaves as
a two level system. In the vicinity of each optimal point, the Hamiltonian
of the system can be written perturbatively as

\begin{equation}
\begin{array}{c}
\hamil=\hamil_{0}-\partial_{\Phi}\left(\alpha E_{J}\cos\left(\frac{\Phi}{\varphi_{0}}-\sum_{j=1}^{3}\varphi_{j,j+1}\right)+\bar{\beta}\frac{1-\bar{d}}{2}\cos\left(\varphi_{23}-\frac{\zeta\Phi}{\varphi_{0}}\right)\right)_{\Phi=\Phi_{opti}}\cdot\left(\Phi-\Phi_{opt}\right)\\
=\hamil_{0}+I_{0}\left[\alpha\text{sin}\left(\varphi_{41}\right)-\zeta\bar{\beta}\frac{1-\bar{d}}{2}\sin\left(\varphi'_{23}\right)\right]\left(\Phi-\Phi_{opt}\right)\equiv\hamil_{0}+\hat{I}\cdot\left(\Phi-\Phi_{opt}\right)
\end{array}\label{eq:pert}
\end{equation}

When the current operator is projected on the eigenstates $\kt 0,\kt 1$
of $\hamil_{0}$ we get

\begin{align}
\left\langle 0\right|\hat{I}\left|0\right\rangle  & =\left\langle 1\right|\hat{I}\left|1\right\rangle \nonumber \\
\left\langle 0\right|\hat{I}\left|1\right\rangle  & =\left\langle 1\right|\hat{I}\left|0\right\rangle \equiv I_{p}
\end{align}

Therefore, the Hamiltonian of the system can be written in this basis
under the form

\begin{equation}
\mathcal{H}_{\text{eff}}=\frac{\hbar}{2}\left[\Delta\sigma_{z}+\varepsilon\sigma_{x}\right]\label{eq: Heff}
\end{equation}

where $\varepsilon=\frac{2I_{p}}{\hbar}\left(\Phi-\Phi_{opti}\right)$.

\section{Flux noise in the tunable flux qubit}

In the following we consider flux noise originating from the SQUID
loop and the remaining loop as two independent noise sources. In order
to estimate their influence on the coherence of the qubits, we will
treat the Hamiltonian perturbatively versus the variables $\phis$
and $\phir=\Phi-\phis$ around an optimal point.

\begin{align}
\hamil & =\hamil_{0}-E_{J}\partial_{\phis}\left(\alpha\cos\left(\frac{\phir+\phis}{\varphi_{0}}-\sum_{j=1}^{3}\varphi_{j,j+1}\right)+\bar{\beta}\frac{1-\bar{d}}{2}\cos\left(\varphi_{23}-\frac{\phis}{\varphi_{0}}\right)\right)_{\Phi=\Phi_{opti}}\delta\phis\nonumber \\
 & -E_{J}\partial_{\phir}\left(\alpha\cos\left(\frac{\phir+\phis}{\varphi_{0}}-\sum_{j=1}^{3}\varphi_{j,j+1}\right)\right)_{\Phi=\Phi_{opti}}\delta\phir\nonumber \\
 & =\hamil_{0}+I_{0}\left(\alpha\sin\left(\varphi_{41}\right)-\bar{\beta}\frac{1-\bar{d}}{2}\sin\left(\varphi'_{23}\right)\right)\delta\phis+I_{0}\alpha\sin\left(\varphi_{41}\right)\delta\phir\nonumber \\
 & \equiv\hamil_{0}+\hat{I}_{S}\delta\phis+\hat{I}_{R}\delta\phir\label{eq:pert2}
\end{align}

One can express the operators $\hat{I}_{S/R}$ in the basis of the
two lowest eigenstates of $\hamil_{0}$ as $\hat{I}_{S/R}=\ixsr\sigma_{x}+\izsr\sigma_{z}+I_{0,S/R}\identity$
and thus write the Hamiltonian as

\begin{equation}
\hamil=\hbar\frac{\Delta}{2}\sigma_{z}+\left(\izs\delta\phis+\izr\delta\phir\right)\sigma_{z}+\left(\ixs\delta\phis+\ixr\delta\phir\right)\sigma_{x}\label{eq:pert2pauli}
\end{equation}

The transition frequency of the qubit can be written as

\begin{align}
\hbar\omega_{01} & \simeq\hbar\Delta+2\izs\delta\phis+2\izr\delta\phir+\frac{1}{2\hbar\Delta}\left(2\ixs\delta\phis+2\ixr\delta\phir\right)^{2}\label{eq:dl2}
\end{align}

Since the flux noise has a 1/f spectrum, one can show that the first
and second order contributions to dephasing are given by \citep{PhysRevB.72.134519}

\begin{align}
\Gamma_{S/R} & =\sqrt{\ln2}A_{S/R}\left|\partial_{\phisr}\omega_{01}\right|\nonumber \\
\Gamma_{2\text{nd},S/R} & \approx2.3A_{S/R}^{2}\partial_{\phisr}^{2}\omega_{01}\label{eq:fluxframwork}
\end{align}

where $A_{S}$ and $A_{R}$ are the flux noise amplitudes in the SQUID
and in the remaining loop respectively.

\begin{align*}
\partial_{\Phi_{S/R}}\omega_{01} & =\frac{2\izsr}{\hbar}+\frac{1}{\hbar^{2}\Delta}\left(2\ixs\delta\phis+2\ixr\delta\phir\right)\left(2\ixsr\right)\\
\partial_{\Phi_{S/R}}^{2}\omega_{01} & =\frac{1}{\hbar^{2}\Delta}\left(2\ixsr\right)^{2}
\end{align*}

An interesting property of 1/f noise is that the decay is Gaussian
for first order contributions. And thus one can write

\begin{equation}
\gffo=\sqrt{\left(\Gamma_{S}\right)^{2}+\left(\Gamma_{R}\right)^{2}}=\sqrt{\ln2\left(A_{S}^{2}\left(\partial_{\phis}\omega_{01}\right)^{2}+A_{R}^{2}\left(\partial_{\phir}\omega_{01}\right)^{2}\right)}\equiv A_{\Phi}\sqrt{\ln2}\left|\partial_{\Phi}\omega_{01}\right|\label{eq:adef}
\end{equation}
where $A_{\Phi}$ is the \emph{apparent} measure of flux noise when
the tunable qubit is biased by an external uniform magnetic field
(see Fig. 1a). 
 Due to the geometrical configuration, $\phis=\zeta\Phi$
and $\phir=\left(1-\zeta\right)\Phi$ and thus

\[
\partial_{\Phi_{S/R}}\omega_{01}=
\underset{\izsr\ll\ixsr}{\underbrace{\cancel{\frac{2\izsr}{\hbar}}}}
+\frac{1}{\hbar^{2}\Delta}\left(2\ixs\zeta+2\ixr\left(1-\zeta\right)\right)\left(2\ixsr\right)\delta\Phi
\]

Since $\hat{I}=\zeta\hat{I}_{S}+\left(1-\zeta\right)\hat{I}_{R}$
(see \ref{eq:pert2}), we have $I_{P}=\zeta\ixs+\left(1-\zeta\right)\ixr$
and thus using \ref{eq:adef}, we get

\begin{align}
A_{\Phi}I_{P} & =\sqrt{A_{S}^{2}\ixs^{2}+A_{R}^{2}\ixr^{2}}
\end{align}

Following Ref \citep{PhysRevApplied.13.054079}, we assume that the
flux noise amplitude is proportional to the square root of the perimeters
$\sqrt{P_{S/R}}$ of the respective sections and define $\gamma\equiv\sqrt{P_{S}/P_{R}}=A_{S}/A_{R}$
such that

\[
\boxed{A_{S}A_{R}=\frac{A_{\Phi}^{2}I_{P}^{2}}{\ixs^{2}\gamma+\ixr^{2}/\gamma}}
\]

Finally we get
\begin{align*}
\gffo & =2\sqrt{\ln2}\frac{\sqrt{A_{S}A_{R}}}{\hbar}\sqrt{\left(\izs+I_{x,S}\frac{2I_{P}}{\hbar\Delta}\delta\Phi\right)^{2}\gamma+\left(I_{z,R}+I_{x,R}\frac{2I_{P}}{\hbar\Delta}\delta\Phi\right)^{2}/\gamma}\\
\gfso & \approx9.2\frac{A_{S}A_{R}}{\hbar^{2}\Delta}\left(\ixs^{2}\gamma+\ixr^{2}/\gamma\right)
\end{align*}

In \ref{fig:fn}b, we represent the calculated contributions of first
and second order flux noise to the dephasing rates for qubits 3 and
4 at optimal point when the flux $\Phi_{S}/\varphi_{0}$ is varied
between $0$ and $2\pi$. The first order contributions are cancelled
when the $\partial_{\phis}\omega_{01}=0$ and reach a maximum of around
200 kHz approximately. The second order contributions are small for
large gap and tend to become dominant when the flux qubit gap is small.

\begin{figure}
\begin{centering}
\includegraphics[width=0.85\linewidth]{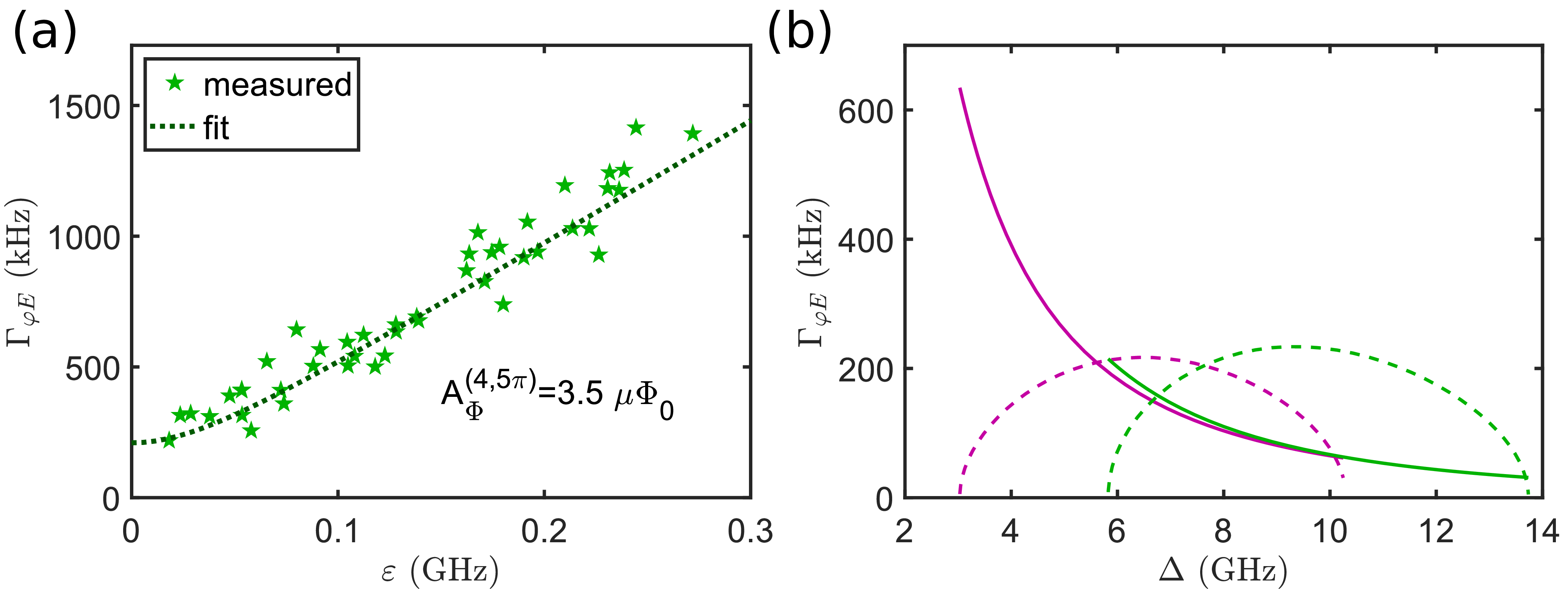}
\par\end{centering}
\caption{\label{fig:fn}(a) Measured pure dephasing rate $\Gamma_{\varphi E}^{\left(4,5\pi\right)}$
versus $\varepsilon$. We extract the apparent flux noise amplitude
from the slope of the graph using \ref{eq:adef}. (b) Calculated contribution
of first (dashed line) and second order (solid line) flux noise to
the dephasing rates for qubits 3 (magenta) and 4 (green) at optimal
point. The curves are calculated for arbitrary values of $\Phi_{S}$
while keeping the qubit at its optimal point.}

\end{figure}

\section{Room temperature measurements of test junctions}

\begin{figure}[H]
\begin{centering}
\includegraphics[width=0.85\linewidth]{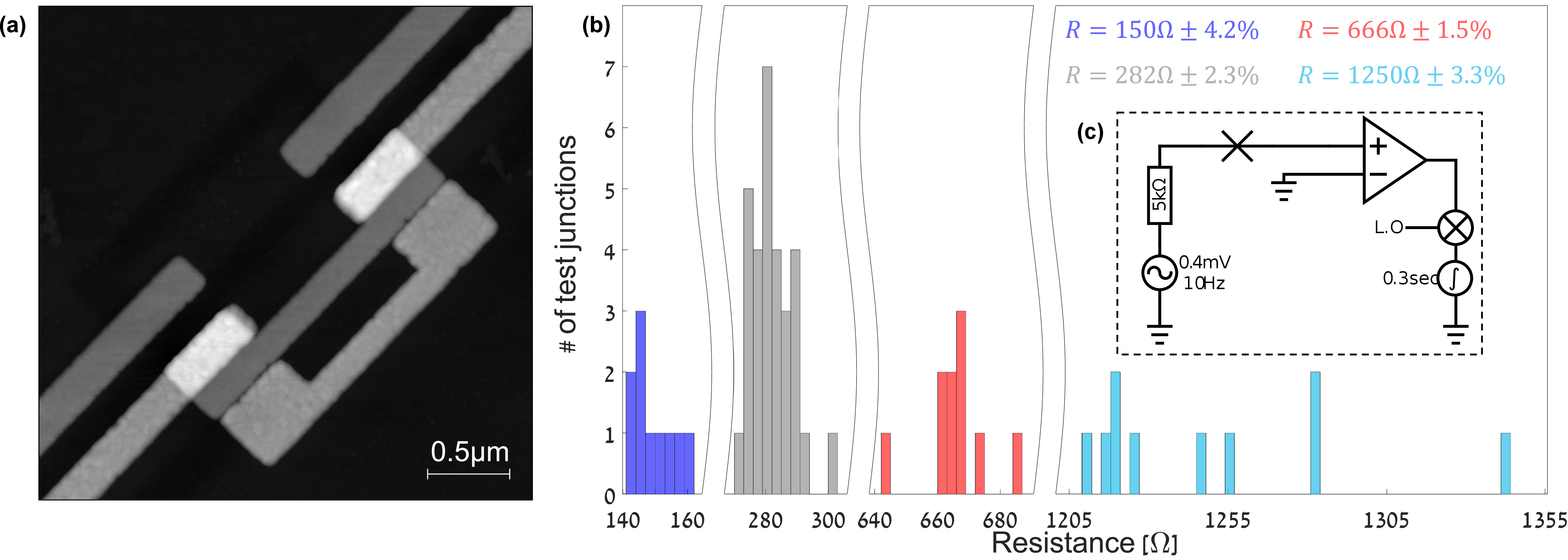}
\par\end{centering}
\caption{(a) Atomic force micrograph of two unitary test junctions in series.
(b) Histogram representing the resistance distribution of an ensemble
of 60 test junctions: 30 unitary junctions (gray), ten $\alpha$ junctions
(red), ten $\beta(1+d)/2$ junctions (dark blue), and ten $\beta(1-d)/2$
junctions (light blue). (c) Scheme of the room temperature lock-in
measurement setup. \label{fig:testjunctions}}
\end{figure}

The IV measurements have been performed using two-probe measurement.
The probe station is connected in series with a $\SI{5}{\kilo\ohm}$
resistor to an A.C. output voltage of amplitude $\SI{4}{\milli\volt}$
at a frequency of $\SI{10.19}{\hertz}$. The current passing through
the junction is measured by a lock-in amplifier with an integration
time of $\SI{300}{\milli\second}$. 

In \ref{fig:testjunctions}b, we present a histogram showing the resistance
distribution of the test junctions. The statistics of the test junctions
presented in \ref{fig:testjunctions}b were taken from 60 test junctions
evaporated simultaneously with the sample. The standard deviations
of the resistances are in the range of $1.5\%\sim4.2\%$. Using these
measurements, we extract the parameters of the qubit $\alpha=0.42\pm0.01$,
$d=0.79\pm0.01$ and $\beta=2.11\pm0.09$. The critical current of
the unitary junction can be estimated by Ambegaokar-Baratoff relation
to be $\SI{1.08}{\micro\ampere}$.

\section{Qubit Parameters}

\begin{table}[H]
\begin{centering}
\setlength{\extrarowheight}{3pt}%
\begin{tabular}{|c|l|c|c|c|c|c|c|}
\cline{3-8} \cline{4-8} \cline{5-8} \cline{6-8} \cline{7-8} \cline{8-8} 
\multicolumn{1}{c}{} &  & $\pi$ & $3\,\pi$ & $5\,\pi$ & $7\,\pi$ & $9\,\pi$ & Units\tabularnewline
\hline 
\multirow{3}{*}{qubit 1} & $\Delta/2\pi$  & \multirow{1}{*}{7.30} & \multirow{1}{*}{12.16} & \multirow{1}{*}{8.15} & \multirow{1}{*}{6.05} & \multirow{1}{*}{11.96} & GHz\tabularnewline
 & $I_{P}$  & \multirow{1}{*}{376} & \multirow{1}{*}{230} & \multirow{1}{*}{342} & \multirow{1}{*}{401} & \multirow{1}{*}{230} & nA\tabularnewline
 & $\chi/2\pi$  & \multirow{1}{*}{5.05} & \multirow{1}{*}{-5.93} & \multirow{1}{*}{6.37} & \multirow{1}{*}{3.16} & \multirow{1}{*}{-6.26} & MHz\tabularnewline
\hline 
\hline 
\multirow{3}{*}{qubit 2} & $\Delta/2\pi$   & 6.84 & 12.82 & 9.82 & 5.60 & 11.91 & GHz\tabularnewline
 & $I_{P}$  & 388 & 238 & 282 & 440 & 263 & nA\tabularnewline
 & $\chi/2\pi$  & 3.61 & -4.31 & 33.1 & 3.41 & -4.31 & MHz\tabularnewline
\hline 
\hline 
\multirow{3}{*}{qubit 3} & $\Delta/2\pi$   & 4.48 & 9.43 & 7.15 & 3.10 & NaN & GHz\tabularnewline
 & $I_{P}$  & 457 & 289 & 354 & NaN & NaN & nA\tabularnewline
 & $\chi/2\pi$  & 2.85 & 16.3 & 5.65 & 3.06 & NaN & MHz\tabularnewline
\hline 
\hline 
\multirow{3}{*}{qubit 4} & $\Delta/2\pi$   & 8.02 & 13.65 & 8.46 & 7.54 & 13.61 & GHz\tabularnewline
 & $I_{P}$  & 365 & 224 & 353 & 341 & 246 & nA\tabularnewline
 & $\chi/2\pi$  & 8.86 & -4.66 & 11.3 & 8.21 & -4.05 & MHz\tabularnewline
\hline 
\hline 
\multirow{3}{*}{qubit 5} & $\Delta/2\pi$   & 6.98 & 12.46 & 9.00 & 5.46 & 11.69 & GHz\tabularnewline
 & $I_{P}$  & 392 & 254 & 332 & 444 & NaN & nA\tabularnewline
 & $\chi/2\pi$  & 2.83 & -4.08 & 6.48 & NaN & -4.23 & MHz\tabularnewline
\hline 
\end{tabular}
\par\end{centering}
\caption{Qubit Parameters: $\Delta$, $I_{P}$ and the dispersive shift of
the resonator $\chi$.}
\end{table}

\begin{center}
\begin{table}[H]
\begin{centering}
\setlength{\extrarowheight}{3pt}%
\begin{tabular}{|c|c|c|c|c|c|c|c|}
\cline{3-8} \cline{4-8} \cline{5-8} \cline{6-8} \cline{7-8} \cline{8-8} 
\multicolumn{1}{c}{} &  & $\pi$ & $3\,\pi$ & $5\,\pi$ & $7\,\pi$ & $9\,\pi$ & Units\tabularnewline
\hline 
\multirow{3}{*}{qubit 1} & $\Gamma_{1}$ & 236 & 1073 & 271 & 296 & 647 & kHz\tabularnewline
 & $\Gamma_{P}$  & 28 & 23 & 77 & 14 & 23 & kHz\tabularnewline
 & $\Gamma_{int}$ & 208 & 1050 & 194 & 282 & 624 & kHz\tabularnewline
\hline 
\hline 
\multirow{3}{*}{qubit 2} & $\Gamma_{1}$ & 194 & 1141 & NaN & 268 & 625 & kHz\tabularnewline
 & $\Gamma_{P}$  & 16 & 34 & NaN & 9 & 47 & kHz\tabularnewline
 & $\Gamma_{int}$ & 178 & 1107 & NaN & 259 & 578 & kHz\tabularnewline
\hline 
\hline 
\multirow{3}{*}{qubit 3} & $\Gamma_{1}$ & 142 & 373 & 185 & 225 & NaN & kHz\tabularnewline
 & $\Gamma_{P}$  & 1 & 151 & 45 & 0 & NaN & kHz\tabularnewline
 & $\Gamma_{int}$ & 141 & 222 & 140 & 225 & NaN & kHz\tabularnewline
\hline 
\hline 
\multirow{3}{*}{qubit 4} & $\Gamma_{1}$ & 446 & 1858 & 341 & 298 & 1568 & kHz\tabularnewline
 & $\Gamma_{P}$  & 79 & 310 & 58 & 62 & 248 & kHz\tabularnewline
 & $\Gamma_{int}$ & 367 & 1548 & 283 & 236 & 1320 & kHz\tabularnewline
\hline 
\hline 
\multirow{3}{*}{qubit 5} & $\Gamma_{1}$ & 236 & 709 & 298 & 236 & NaN & kHz\tabularnewline
 & $\Gamma_{P}$  & 14 & 8 & 60 & 5 & NaN & kHz\tabularnewline
 & $\Gamma_{int}$ & 222 & 701 & 238 & 231 & NaN & kHz\tabularnewline
\hline 
\end{tabular}
\par\end{centering}
\caption{Relaxation Rates: measured relaxation rate $\Gamma_{1}$, estimated
Purcell rate $\Gamma_{P}$, and the estimated intrinsic rate $\Gamma_{int}=\Gamma_{1}-\Gamma_{P}$.}
\end{table}
\par\end{center}
\end{widetext}

\bibliographystyle{unsrt}
\bibliography{references_tunablequbits}

\end{document}